\documentclass[]{spie}  

\usepackage{amsmath,amsfonts,amssymb}
\usepackage{graphicx}
\usepackage[colorlinks=true, allcolors=blue]{hyperref}

\title{A Federated Capability-based Access Control Mechanism for Internet of Things (IoTs)}

\author[a]{Ronghua Xu}
\author[a]{Yu Chen*}
\author[b]{Erik Blasch}
\author[c]{Genshe Chen}

\affil[a]{Dept. of Electrical \& Computer Engineering, Binghamton University, Binghamton, NY 13902}
\affil[b]{U.S. Air Force Office of Scientific Research (AFOSR), Arlington, VA 22203}
\affil[c]{Intelligent Fusion Technology, Inc. Germantown, MD 20876}

\authorinfo{Further author information: (Send correspondence to Yu Chen.)\\Yu Chen.: E-mail: ychen@binghamton.edu, Telephone: 1 607 777 6133}

\begin{document}
\maketitle

\begin{abstract}
The prevalence of Internet of Things (IoTs) allows heterogeneous embedded smart devices to collaboratively provide intelligent services with or without human intervention. While leveraging the large-scale IoT-based applications like Smart Gird and Smart Cities, IoT also incurs more concerns on privacy and security. Among the top security challenges that IoTs face is that access authorization is critical in resource and information protection over IoTs. Traditional access control approaches, like Access Control Lists (ACL), Role-based Access Control (RBAC) and Attribute-based Access Control (ABAC), are not able to provide a scalable, manageable and efficient mechanisms to meet requirement of IoT systems. The extraordinary large number of nodes, heterogeneity as well as dynamicity, necessitate more fine-grained, lightweight mechanisms for IoT devices. In this paper, a federated capability-based access control (FedCAC) framework is proposed to enable an effective access control processes to devices, services and information in large scale IoT systems. The federated capability delegation mechanism, based on a propagation tree, is illustrated for access permission propagation. An identity-based capability token management strategy is presented, which involves registering, propagation and revocation of the access authorization. Through delegating centralized authorization decision-making policy to local domain delegator, the access authorization process is locally conducted on the service provider that integrates situational awareness (SAW) and customized contextual conditions. Implemented and tested on both resources-constrained devices, like smart sensors and Raspberry PI, and non-resource-constrained devices, like laptops and smart phones, our experimental results demonstrate the feasibility of the proposed FedCAC approach to offer a scalable, lightweight and fine-grained access control solution to IoT systems connected to a system network.
\end{abstract}


\section{INTRODUCTION}
\label{sec:intro}  

With the proliferation of the Internet of Things (IoTs), the number of physical devices are being connected to the Internet at an unprecedented scale. The prevalence of the IoTs changes human activities by ubiquitously providing applications and services revolutionizing transportation, healthcare, industrial automation, and emergency response \cite{al2015internet}. These capabilities offer both situation awareness and measurement data to provide context \cite{snidaro2016context, blasch2017panel}.

While benefiting from the large-scale applications like Smart Gird and Smart Cities, future IoT systems also incur more concerns on security and privacy. With the increased popularity, the connected smart IoT devices with insufficient security enforcement increases the risk of privacy breaches and various attacks. Security issues, such as privacy, authentication, access control, system configuration, information storage and management, are the main challenges facing any IoT environment \cite{alaba2017internet}.

Among the top security challenges that IoTs face, access authorization is critical in resource and information protection. Conventional access control approaches, like Access Control List(ACL), Role-based Access Control(RBAC) and Attribute-based Access Control(ABAC) have been widely used on IT system. However, they are not able to provide a scalable, manageable and efficient mechanism to meet requirements of IoT networks:

\begin{itemize}
\item \textit{Scalability}: The fast growing number of devices and services also pose increasing management overload in access control systems that are based on ACL or RBAC models. Access control strategies are expected to be able to handle the scalability problem resulting from the large scale IoT networks.
\item \textit{Heterogeneity}: IoT system normally integrate heterogeneous cyber physical objects with variant underlying technologies or in different application domains, and each domain or platform has its own specific requirements for identity authentication and authorization policy enforcement. Both RBAC and ABAC have been found inflexible to provide complex arrangements to support delegation and transitivity, which are essential for efficient and effective intra-domain authorization and access control.
\item \textit{Causality}: Traditional RBAC and ABAC systems envisage planned and long-lived patterns, while the IoT world is mainly characterized by short-lived, often causal and/or spontaneous interactions \cite{gusmeroli2013capability}, in which access control scheme is required to deal with dynamic challenges.
\item \textit{Lightweight}: IoT devices are usually resource-constrained, which cannot support heavy computational and large storage required applications, and those smart devices connect to each other by low power and lossy networks. Consequently the access control protocol should be lightweight and does not impose significant overhead on devices and communication networks.
\end{itemize}

The extraordinary large number of devices with heterogeneity and dynamicity necessitate more scalable, flexible and lightweight access control mechanism for IoT networks. In this paper, we propose a federated capability-based access control(FedCAC) framework, which provides a scalable, fine-grained and lightweight access control solution to protect devices, services and information in IoT networks. An \textit{identity-based capability token management strategy} is presented and the authorization delegation mechanism is illustrated. A capability-based access validation process is implemented on service providers that integrate situational awareness (SAW) and customized contextualized conditions. Experimental results demonstrate the feasibility and effectiveness of the proposed FedCAC scheme. In summary, the major contributions are:

\begin{itemize}
\item[1)] A complete architecture of federate capability-based authorization system is proposed, which includes capability management, delegation authority, and access right validation.
\item[2)] A concept-proof prototype is implemented and tested on both resource-constrained devices and non-resource-constrained devices, and the experimental results validate the feasibility of the FedCAC scheme in IoT environments without introducing significant overhead.
\end{itemize}

The remainder of this paper is organized as follows: Section \ref{sec:related} analyzes and reviews the state of the art research in access control for IoT systems. Section \ref{sec:FedCAC} illustrates the details of the proposed federated capability based access control system. Section \ref{sec:proto} explains the implementation of the proof-of-concept prototype. The experimental results and evaluation are discussed in Section \ref{sec:exp}. Finally, the summary, current limitations and prospective solutions are discussed in Section \ref{sec:con}.

\section{Related work}
\label{sec:related}  

Authentication and access control technologies are among the main elements to protect the security and privacy in IoT devices \cite{ouaddah2017access}. It's critical to prevent unauthorized subjects from having access the IoT device, service or information. As a fundamental mechanism to enable security in computer systems, \textit{access control} is the process that decides who is authorized to have what communication rights on which objects with respect to some security models and policies \cite{gong1989secure}. An effective access control system should satisfy the main security requirements of confidentiality (no unauthorized disclosure of resources), integrity (no improper modifications of resources), and availability (ensuring accessibility of resources to legitimate users) \cite{suhendra2011survey}.  Newly raised security and privacy issues in the era of IoTs require that access control systems should be built on principals of high scalability, flexibility, lightweight and causality.

Recently, various access control methods and solutions with different objectives have been proposed to address security challenges in IoTs. The Role-Based Access Control (RBAC) model \cite{sandhu1996role} provides a framework that specifies user access authorization to resources based on roles, and supports principals such as least privilege, partition of administrative functions and separation of duties \cite{samarati2000access}. However, a pure RBAC model presents a role explosion problem when the amount of resources grows or the access control policies cover many administrative domains \cite{sandhu1996role}, which is inappropriate to implement security policies that require interpreting complex and ambiguous IoT scenarios. RBAC model implemented on IoTs adopts a service-based approach \cite{de2008socrades,spiess2009soa}, where each IoT device offers its functionality as a standard web service. An authorization request is verified by access control process at service provider before being performed by the service application. The RBAC model was extended by introducing context constraints to consider contextual awareness in access control decisions \cite{zhang2010extended}. However, the mapping mechanism between physical objects and web services is not clearly described and the smart objects in authorization decision-making is not considered. Similar proposals are presented by integrating RBAC model into the Web of Things (WoTs) approach to implement access control policies on the smart objects via the web service \cite{jindou2012access, barka2015securing}. However, those proposals are not able to clearly specify the fine-grained access control on variant resources, like the mapping of the role notion and device-to-device communication.

To address the weaknesses of RBAC model in a highly distributed network environment, Attribute-based Access Control (ABAC) \cite{yuan2005attributed, smari2014extended} is introduced in IoT networks to reduce the number of rules resulted from role explosion. In ABAC access control policies are defined through directly associating attributes with subjects. An efficient authentication and ABAC based authorization scheme for the perception layer of IoTs have been proposed \cite{ye2014efficient}. Mutual authentication is implemented by elliptic-curve cryptography (ECC) based secure key establishment such that it incurs much lower storage and communication overhead on the resource constrained devices running in the IoT perception layer. Based on user attribute certificates, an access right is granted by access control authority to ensure fine-grained access control. However, specifying a consistent definition of the attributes within a domain or across different domains could significantly increase effort and complexity on policy management as the number of devices grow, and hence, the proposal is not suitable on large scale distributed IoT networks.

Due to drawbacks that exist in traditional access control models such as RBAC and ABAC, the requirements imposed by IoT scenarios cannot be satisfied. Given many great advantages from an IoT perspective, such as scalability, flexibility, distributed, user-driven, IoT systems can support delegation and revocation \cite{ouaddah2017access}. Capability-based access control approaches have been considered a promising solution to IoTs. The Access Control Matrix (ACM) model represents a good conceptualization of authorizations by providing a framework for describing discretionary access control (DAC) \cite{samarati2000access}. As two implementations of access matrix, Access Control List (ACL) and Capability are widely used in authorization system. Detailed comparisons between ACL and capability-based access control were reported in earlier publications \cite{close2009acls, sandhu1992typed}.

In the ACL model, each object is associated with an access control list that saves the subjects and their access rights for the object. The ACL is a centralized approach by nature so that it supports administrative activities with better traceability by implementing access control strategy on cloud servers \cite{liu2014information}. However, as the number of subjects and resources increases, confused duty problems are identified in ACL and access rules become much more complex to manage. Due to the centralized management property, ACL cannot provide multiple levels of granularity, is not scalable and is vulnerable to single point of failure. Meanwhile, in capability model each subject is associated with a capability list that represents its access rights to all concerned objects. The concept of capability was first introduced in 1996 and defined as a token, ticket, or key that gives the possessor permission to access an entity or object in a computer system\cite{dennis1966programming}. Unlike the ACL model that relies on centralized access control policy, in Capability-based Access Control (CapAC) model, capability token validity and access right authorization process could be executed by local service providers. Through enforcing authorization processes locally on distributed edge devices, device-to-device communication is feasible and complexity of centralized policy management is reduced. CapAC has been implemented in many large scale IoT-based projects, like IoT@Work\cite{gusmeroli2012iot}.

Although capability-based methods have been used as a feature in many access control solutions for the IoTs applications, applying the original concept of capability-based access control model in IoT network has raised several issues. There is a detailed discussion about the main characteristics and a comparison between ACL and capability, and two major drawbacks in the original capability access control model are identified: capability propagation and revocation~\cite{gong1989secure}. To tackle these challenges, a Secure Identity-Based Capability (ICAP) System is proposed, which aimed at a distributed system in a networked environment. The proposed ICAP model enables the monitoring, mediating, and recording of capability propagations to enforce security policies as well as achieving rapid revocation capability by using an exception list. However, the centralized access control server (ACS) is the bottleneck of system performance and becomes more vulnerable to denial of service (DoS) attacks. The author didn't provide a clear illustration on security policy used in capability generation and propagation, neither was the context information in making authorization decision considered \cite{gong1989secure}.

Based on ICAP an approach that integrated authentication and access control was proposed, which is called the Identity Authentication and Capability-based Access Control (IACAC) model~\cite{mahalle2013identity}. The IACAC solution presented many suitable merits for IoTs, like low computing time, fine granularity delegation and scalability; however, capability token propagation and revocation was not addressed. To enable context awareness in federated IoTs, an authorization delegation method was proposed based on Capability-based Context-Aware Access Control (CCAAC) model~\cite{anggorojati2012capability}. By introducing delegation mechanism to capability generation and propagation process, the CCAAC model shows great advantages to address scalability and heterogeneity issues in IoT networks. Given the requirement that a prior knowledge of the trust relationship among domains in federated IoTs must be established, however, the proposed approach is not suitable for all IoT scenarios. Inspired by the SUN DIGITAL ECOSYSTEM ENVIRONMENT project \cite{skinner2009cyber}, a \textit{capability-based access control (CapAC)} model was proposed that adopted a centralized approach for managing access control policy \cite{gusmeroli2013capability}. Access right delegation allows a more sophisticated access control customization and the revocation service framework is integrated in the system. The proposed CapAC could effectively support the flexibility, dynamicity and scaling needs in IoT contexts. However, the proposal didn't consider lightweight requirement at the smart device side. The proposed access control scheme is not suitable to be deployed on IoT devices. To address limitation in CapAC, a Distributed Capability-based Access Control(DCapAC) model was proposed, which was directly deployed on resource-constrained devices~\cite{hernandez2013distributed, hernandez2016dcapbac}. Meanwhile, DCapAC was extended to a flexibility trust-aware access control system for IoTs (TACIoT)~\cite{bernabe2016taciot}. The DCapAC allows smart devices to autonomously make decisions on access rights, based on authorization policy, and it shows advantages in scalability and interoperability. However, capability revocation management and delegation were not discussed, neither were the granularity and context-awareness considered.

Unlike the approaches mentioned above, our proposal, a \textit{Federated Capability-based Access Control model (FedCAC)}, could effectively deal with scalability, granularity, and dynamicity challenges in access control strategy for IoTs. Through delegating part of the identify authentication and authorization task to domain delegator, workload of the centralized policy decision making center (PDC) is reduced. Processing validation of capability on local service providers enables a lightweight, context-aware and fine-grained access control mechanism for IoT devices. Involving smart objects in access right authorization process allows device-to-device communication, which implies a more scalable, distributed and inter-operable IoT network environment.

\section{Federated Capability-based Access Control System}
\label{sec:FedCAC}

In most IoT based systems, data processing and security enforcement are deployed on centralized cloud center where abundant computing and storage resources are allocated. As a result, all access right requests from devices need to be transmitted to remote servers for authentication and authorization. Such a centralized network architecture is not scalable for today's large IoT networks, and latencies are not tolerable in many mission-critical applications. To meet the requirements of real-time processing and instant decision making,  online, uninterrupted smart surveillance systems have been intensively studied recently leveraging the edge-fog-cloud computing paradigm \cite{chen2016dynamic, chen2017enabling, chen2016smart}. Inspired by the proposed automatic surveillance system architecture, a federated capability-based access control system (FedCAC) is proposed in this paper, which consists of three hierarchical layers:

\begin{itemize}
\item \textit{Cloud Computing Layer}: This layer handles global management tasks that require high computational power and large storage space. Global profile database and policy decision making center (PDC) are deployed at the cloud server, which are responsible for identity and policy rules management.
\item \textit{Fog Computing Layer}: Federated delegation mechanism is implemented on near-site fog computing nodes, such as laptops, tablets or smart phones, which is called \textit{coordinator}. In each application-specific IoT network, an IoT coordinator is chosen by PDC among smart devices in domain network, and delegates domain-specific access authorization policies and identity management tasks from the PDC.
\item \textit{Edge Computing Layer}: Access control authorization is processed at these on-site edge computing devices, which are responsible for performing requested tasks as service providers.
\end{itemize}

According to the above proposed hierarchical architecture, we have implemented a FedCAC framework in a physical IoT network environment to verify the efficiency and effectiveness. The next subsection provides a comprehensive system design of FedCAC framework.

\subsection{System Design of FedCAC}
Figure \ref{fig:1-FCapAC} illustrates the proposed FedCAC system architecture. The FedCAC covers two IoT-based service request scenarios: intra-domain and inter-domain. Operation and communication modes are listed as:

\begin{figure} [ht]
\begin{center}
\begin{tabular}{c}
\includegraphics[height=9cm]{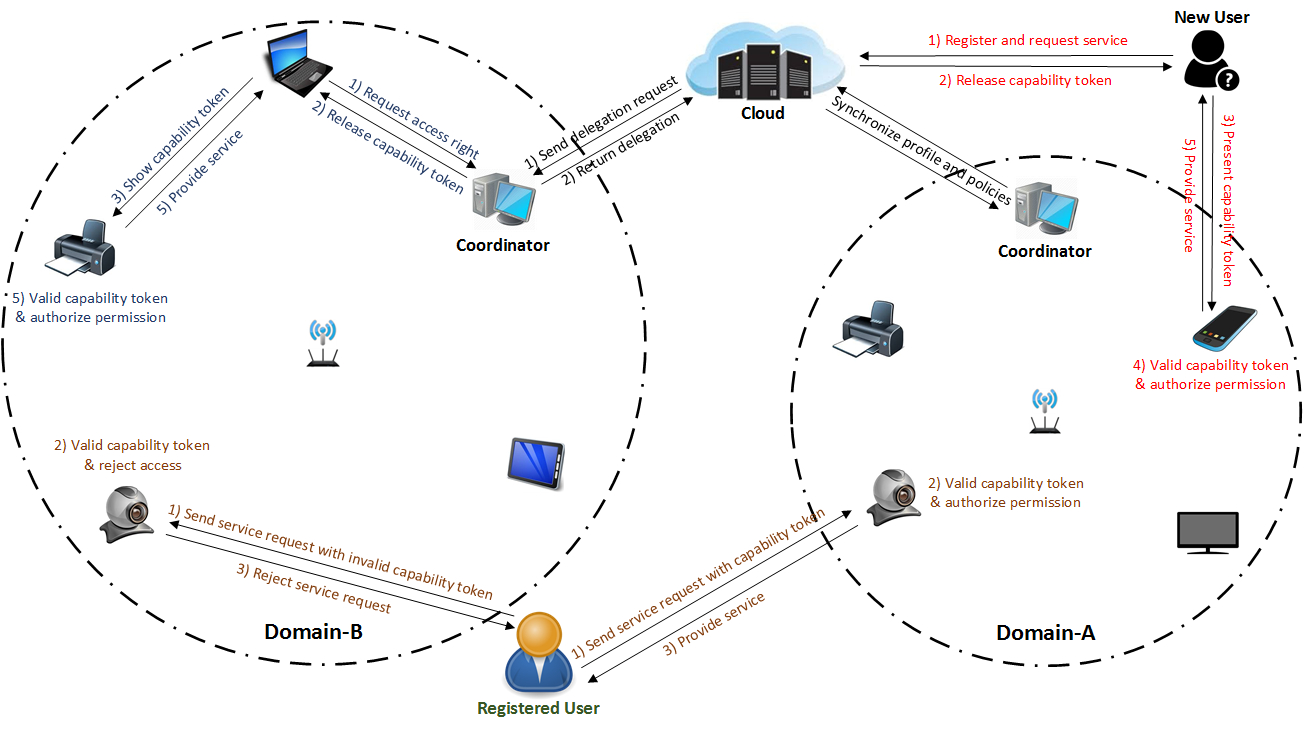}
\end{tabular}
\end{center}
\caption[example] { \label{fig:1-FCapAC} System Architecture of FedCAC.}
\end{figure}

\begin{itemize}
\item \emph{Registration}: All objects must be registered to a global profile database that provides device identity authentication and management. Once the identity information related to users or IoT devices is verified, the profile of each registered entity is created with a globally unique virtual device ID for authentication process when access right request happens. Only registered entities are allowed to join the domain network for further operations, such as service requesting and providing.
\item \emph{Delegation}: The PDC at cloud server is responsible for policy definition and access right authorization enforcement. To reduce the overhead of the centralized cloud server and meet requirements of scalability and heterogeneity in each IoT domain, the domain coordinator delegates part of the policy decision making tasks and carries out domain specified authorization rules based on domain specified policies. After a federated delegation relationship has been established between the cloud server and the coordinator, the cloud server periodically synchronizes profile and policy data with the coordinator.
\item \emph{Capability Propagation}: To successfully access services or resources at service providers, an object initially sends access right request to the domain coordinator or cloud server to get capability token. Given registered object information established in profile database, a policy decision making module evaluates the access request by enforcing defined authorization policy rules. If the access request is granted, the capability token encapsulating access right is generated with issuer's signature and sent back to the object. Otherwise, the access right request is rejected. Then the object requests access to resources by presenting a valid capability token to the service provider.
\item \emph{Authorization Validation}: Given the assumption that the device identity authentication has been finished and a security communication channel has been established between the service requester and provider, an authorization validation process is performed at local service providers on receiving a service request from the object. Considering capability token validation and an access authorization process result, if the access right policies and conditional constraints are satisfied, then the service provider grants the access request and offers services to the requester. Otherwise, the service request is denied.
\end{itemize}

To enable a scalable, distributed and fine-grained access control solution to IoT networks, the proposed FedCAC is focused on three issues: identity-based capability management, access right authorization and privilege mechanism delegation.

\subsection{Capability Token Structure}
The proposed FedCAC model shares the capability token structure from the identity-based capability (ICAP) scheme \cite{anggorojati2012capability}. In our FedCAC system, entities are categorized as subjects and objects. \textit{Subjects} are defined as entities who request service from service providers, while \textit{objects} are referred to entities who offer resources or services. Entities could be either human beings or smart devices. In the global profile database, all registered entities are associated with a global unique Virtual Identity (VID) which is used as the prime key for identifying entities' profile information. The VID is essentially a hash value generated by applying hash function to a profile record and signed by an object owner. The VID is defined as follow:

\begin{equation}
\label{eq:VID}
VID = f( P, Sign)
\end{equation}
\begin{itemize}
\item $f$ : one-way hash function;
\item $P$ : profile that includes entities' attributes and personal information;
\item $Sign$ : signature signed by entity's owner.
\end{itemize}

Smart objects are service providers, and each object specifies a set of access rights that can be granted to a service requester. The meta data of access rights are defined as an internal capability:

\begin{equation}
\label{eq:inCap}
_{in}Cap_{0} = \{ VID_{O}, AR, Rnd_{0} \}
\end{equation}
where
\begin{equation}
\label{eq:Rnd0}
Rnd_{0} =  f( VID_{O}, AR )
\end{equation}
\begin{itemize}
\item $f$ : one-way hash function
\item $VID_{O}$ : virtual ID of object that provides service or resource;
\item $AR$ : set of access rights for action, e.g. read, write, execute;
\item $Rnd$ : random number generated by a one-way hash function to prevent forgery.
\end{itemize}

All internal capability tokens are stored and managed by the centralized PDC. Once the delegation relationship between the domain coordinator and the PDC has been established, those $_{in}Cap_{0}$ that are related to objects in the domain will be copied to the domain coordinator. When the $i$-th subject ($S_i$) sends service access request to the PDC or the coordinator for accessing service on an object ($O$), then the external capability is generated by encapsulating identity information. In general, the external capability specifies which subject can access resources of a target object by associating subject, object, actions and condition constraints. The identity-based external capability structure is defined as follow:
\begin{equation}
\label{eq:extICap}
_{ext}ICap_{i} = \{ VID_{S}, VID_{O}, AR, C, Rnd_{i}\}
\end{equation}
where
\begin{equation}
\label{eq:Rndi}
Rnd_{i} =  f( VID_{S}, VID_{O}, AR, C, Rnd_{0} )
\end{equation}

\begin{itemize}
\item $f$ : one-way hash function;
\item $VID_{S}$ : virtual ID of subject that requests access to service or resource;
\item $VID_{O}$ : virtual ID of object that provides service or resource;
\item $AR$ : set of access right for action, e.g. read, write, execute;
\item $C$ : set of context awareness information, such as time, location; and
\item $Rnd$ : random number generated by a one-way hash function to prevent forgery.
\end{itemize}

In our FedCAC access control system, AR is defined as the access right set. For example, the AR can be \{Read\}, \{Write\}, \{Read; Write\}, or \{NULL\}. If AR=\{NULL\}, the operation conducted on the resource is not allowed. While $C$ is defined as a context constraints set, like $C={C1, C2}$ or $C={NULL}$. If $C={NULL}$, no context constraint is considered in the access right validation process.

\subsection{Capability-based Access Right Authorization}
\begin{figure} [ht]
\begin{center}
\begin{tabular}{c}
\includegraphics[height=9.3cm]{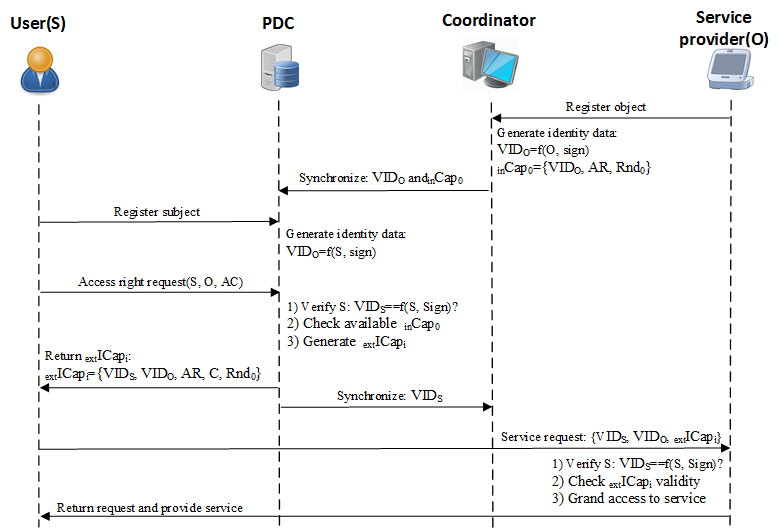}
\end{tabular}
\end{center}
\caption[example] { \label{fig:2-CapAuthorization} Flowchart of the Capability-based access right authorization.}
\end{figure}

Given the defined FedCAC structure, access right authorization is implemented on the PDC or the domain coordinator by signing external capability tokens and distributing to subjects. As shown by Fig. \ref{fig:2-CapAuthorization}, a comprehensive capability-based access right authorization procedure consists of identity registration, capability generation, access authorization and access right validation.

\begin{itemize}
\item \emph{Identity registration}: Given all subjects and objects have been documented in the profile database after the registration process, virtual IDs ($VID_S$ and $VID_O$) are generated by applying an hash function to the profile data and simply signed by the owner to prevent identity spoof. VIDs can be generated either by the PDC or the domain coordinator given the assumption that mutual authentication and secure communication establishment have been achieved.
\item \emph{Capability generation}: As a type of meta data to represent access right, internal capability $_{in}Cap_{0}$ could integrate identity information into a basic capability by associating VID with AR, thus the $_{in}Cap_{0}$ has the identified property to prevent forgery. Large number of $_{in}Cap_{0}$ are grouped into the capability pools on a PDC cloud server, and the capability pools data could be synchronized among the PDC and coordinators periodically to enable data consistence.
\item \emph{Access authorization}: When user sends a request to access a service or resource on service provider, the access authorization procedure is performed by the PDC or a coordinator based on the predefined access control policy. Then an external capability $_{ext}ICap_{i}$ is generated and distributed to the user if the access request is authorized. The external capability $_{ext}ICap_{i}$ is one type of token or key for a user to present as a certificate to access a service or resource. The $_{ext}ICap_{i}$ is generated by associating $_{in}Cap_{0}$ with $VID_S$ and local context constraints, allowing the authorized subject to perform granted operations under specific conditions, such as certain locations or time duration.
\item \emph{Access right validation}: As the service provider received a request from a subject, it makes a decision on whether or not to grant access to the service according to the local access control policy. Implementing access right validation at the local service provider allows smart objects to be involved in the access control decision making task, which is suitable to offer a flexible and fine-grained access control service in IoT networks.
\item \emph{Capability revocation}: Capability revocation considers two scenarios:  $_{in}Cap_{0}$ revocation and $_{ext}ICap_{i}$ revocation. The $_{in}Cap_{0}$ revocation may result from access control policy modification or the target service becomes unavailable. In this case, the PDC just simply removes revoked $_{in}Cap_{0}$, then service providers deny all requests with access right encapsulated in $_{in}Cap_{0}$. In case of $_{ext}ICap_{i}$ revocation, which withdraws authorized access right kept by specific subject, the PDC or coordinator could generate a capability revocation certificate that includes the revoked subject VID signed by the issuer's signature, and distributes it to the service provider. The service provider maintains a capability revocation list containing revoked identities' information, such as VID and expire time. Any request from entities in the capability revocation list will be denied the ability to grant access to services.
\end{itemize}

In the proposed capability-based access right authorization procedure, both the PDC and coordinator are responsible for identity registration and access right authorization. The major difference between the PDC and coordinator is that the PDC holds the root authority of global identity management and authorization decision making, while the coordinator delegates part of privileges in the registration and decision making task to reduce the workload of the centralized PDC server and meet the requirement of the highly heterogeneous IoT environment. Our FedCAC framework is realized by introducing delegation mechanism to capability-based access control.

\subsection{Delegation Mechanism}
\begin{figure} [ht]
\begin{center}
\begin{tabular}{c}
\includegraphics[height=9.3cm]{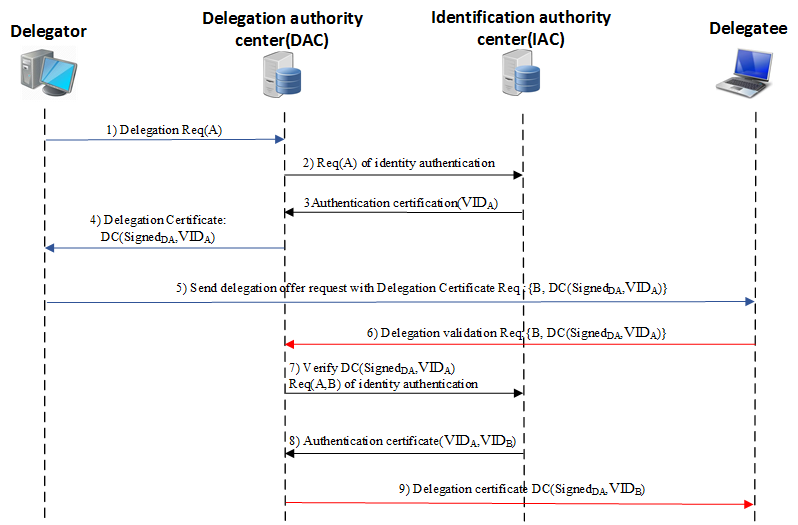}
\end{tabular}
\end{center}
\caption[example] { \label{fig:3-Delegation} Delegation process in FedCAC system.}
\end{figure}

Delegation enables an entity to give permission to let other entities function on its behalf by providing all or some of its rights. It is considered a useful and effective approach to improve the scalability of distributed systems and decentralize access control tasks \cite{gomi2005delegation}. Our delegation mechanism is inspired by an identity delegation framework \cite{gomi2011dynamic}, which adopted the access tokens across security domains in a federated network environment. Figure \ref{fig:3-Delegation} illustrates the delegation process in our proposed FedCAC system.

In FedCAC system, the delegator, delegation authority center (DAC) and identification authority center (IAC) are all implemented as service applications on the cloud server, while the delegatee is deployed on the coordinator in each network domain. The DAC is responsible for generating a Delegation Certificate (DC) and the IAC takes care of identification management and authentication service. Prior to the delegation process, the delegator and delegatee should have established a trust relationship through an identity authentication process. The involved work flow in delegation process is:

\begin{itemize}
\item \emph{Delegation request}: The delegator sends a delegation request to the DAC to ask for Delegation Certification (DC). On receiving the request from the delegator, the DAC verifies the identity of delegation by sending identify authentication request to the IAC. If the delegator's identity is valid, the IAC returns a virtual ID (VID) of the delegator to the DAC. Since the root delegator is on the server, the VID is the hash value of a cloud identity information, like a public key. Then the DAC signs the VID by using the cloud identity's private key and sends back the singed DC to the delegator. Otherwise, the DAC rejects the delegation request by returning a failure notification.

\item \emph{Delegation propagation}: After fetching the DC from the DAC, the delegator is capable of propagating delegation to any trusted delegatee by sending delegation offering request with the DC. To simplify the delegation chain management, the delegation propagation depth in our delegation mechanism is limited to 1 level, which means that only the root delegator (cloud server) has the privilege of delegation propagation. The root delegator on the cloud server has the authority to choose coordinator who acts as a delegatee in each network domain, while a coordinator only performs delegated privileges but cannot transfer its delegated access rights further to other entities.

\item \emph{Delegation acknowledgement}: On receiving a delegation offer request from the delegator, the delegatee (coordinator) sends a delegation authentication request to the DAC to validate the delegation offer. After received the request from delegatee, the DAC launches identity authentication request to the IAC to validate the identity information of the delegator and delegate. If identity authentication is successful, an authentication certificate including the VID of the delegator and delegatee is sent back to the DAC. Finally, the DAC generates a new DC by signing delegatee's VID and distributes it to the coordinator.

\item \emph{Delegation revocation}: Since our system only supports one level depth delegation propagation, the DAC maintains a global delegation revocation list (DRL), which contains the VIDs of the entities whose delegation rights have been revoked. In the revocation process, the trust relationship between the old coordinator and the DAC is nullified, then the DAC signs a revocation certificate including the revoked coordinator's VID and sends it to the new delegatee (coordinator). The new coordinator broadcasts the revocation certificate to all nodes in the domain to nullify the capability tokens issued by the revoked coordinator. Once the revocation process is conducted successfully, each node only maintains the latest revocation certificate.
\end{itemize}

\section{Prototype Design}
\label{sec:proto}
The proposed FedCAC approach is implemented as an web service application based on the Flask framework \cite{flash} using Python. Flask is a micro-framework for Python based on Werkzeug, Jinja 2 and good intentions. Lightweight and extensible micro architectures enable the Flask a preferable web solution on resource constrained IoT devices.

\subsection{The FedCAC Prototype Implementation}
A prototype of FedCAC system has been implemented. It consists of two main parts: a client and a server. The client is the service consumer who sends a service request to server, which is the service provider who offers REST-ful applications programming interface (API) for the client to obtain data or perform operations on resources at the server side. The identity authentication process runs on both the client and server, and the trust and security communication is established before accessing resources.

\subsubsection{Capability Token Structure}
To deploy a lightweight access control system on resource constrained platforms, like smart sensors and Raspberry PI, the FedCAC chooses JavaScript Object Notation (JSON) as the representation format of the capability token. Figure \ref{fig:4-Cap_token} demonstrates a capability token example used in FedCAC. The data fields in the capability token structure are described as follow:

\begin{figure} [ht]
\begin{center}
\begin{tabular}{c}
\includegraphics[height=8cm]{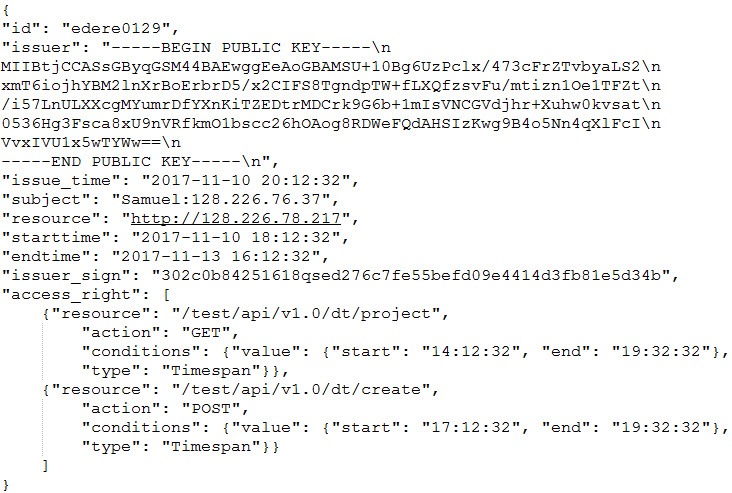}
\end{tabular}
\end{center}
\caption[example] { \label{fig:4-Cap_token} Capability token data structure.}
\end{figure}

\begin{itemize}
\item $id$ : This field is used to unequivocally identify a capability token;
\item $issuer$: The entity who issues the token and signs it using issuer's private key. In our case, we use issuer's public key infrastructure (PKI) to uniquely identify the issuer;
\item $issue\_time$: It is used for recognizing the time when the token was issued;
\item $issue\_sign$: This field carries the digital signature of the token;
\item $subject$: It refers to the identity to which the rights from the token are granted;
\item $resource$: It usually refers to the address of service provider to which the token applies;
\item $starttime$: The time when token becomes active;
\item $endtime$: The time when token becomes expired;
\item $access\_right$: This field defines a set of rights that the issuer has granted to the subject, including
	\begin{itemize}
	\item $action$: It is used to identify a specific granted operation over resource;
	\item $resource$: It represents the resource at the service provider for which the operation is granted. In our case, resource is defined as granted REST-ful API;
	\item $conditions$: It represents a set of conditions that must be fulfilled locally on the service provider to grant the corresponding operation.
	\end{itemize}
\end{itemize}

\subsubsection{Access Authorization Service}
\begin{figure} [ht]
\begin{center}
\begin{tabular}{c}
\includegraphics[height=8cm]{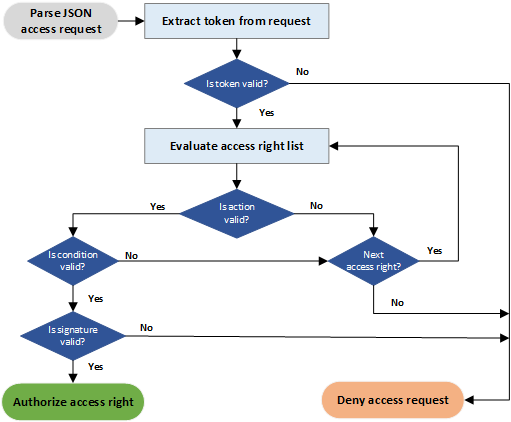}
\end{tabular}
\end{center}
\caption[example] { \label{fig:5-Fig5_CapAC_processing} Capability-based access authorization process.}
\end{figure}

The capability based access control scheme is deployed at the server side by intercepting the API request to enforce access right authorization policies on local service provider. The access right validation process is launched after an API request containing the capability token has been received by the server. Figure \ref{fig:5-Fig5_CapAC_processing} shows a block diagram that illustrates the whole access right validation process.

\begin{itemize}
\item \emph{Token validation}: After a capability token is extracted from a request as JSON data by the service provider, the token is first processed by checking its $issue\_time$, $starttime$ and $endtime$. In the case that the token is not valid, the authorization process stops and sends a deny access request acknowledgement to the subject.

\item \emph{Action Grant Checking}: The service provider checks each element in the access right set to make sure that the requested operation is permitted. The process checks whether or not the REST-ful method used by the requester matches the action filed in a specific access right element and the value of resource field is the same as Request-URI  (uniform resource identifier) option in the request message. If one of the element verifications fails, the process takes the next access element for evaluation. If none of the elements pass verification, the authorization process stops and denies the access request.

\item \emph{Condition Verification}: Once the action on a target resource is permitted, the conditions are evaluated on the local device to verify whether the conditions of the token are satisfied. The process goes through each element in the condition set to find the matched one. If no condition is fulfilled in the local environment, the authorization process stops and denies access request.

\item \emph{Token Signature Checking}: Finally, the signature of the token has to be verified before granting the permission for operations on the resource. The validation process uses issuer's public key stored in issuer filed to validate the token signature. Since a digital signature validation usually needs cryptographic operations that require a lot of computational resources, the token signature verification is left to the final step.
\end{itemize}
	
\section{Experimental results}
\label{sec:exp}
We have validated the feasibility of our proposed FedCAC scheme by deploying a concept-proof prototype both on fog computing nodes and a cloud computing server. The experimental setup and results are discussed in this section.

\subsection{Testbed Setup}
The fog computing node is a Raspberry PI 3 Model B with the configuration as follows: 1.2GHz 64-bit quad-core ARMv8 CPU, the memory is 1GB LPDDR2-900 SDRAM and the operation system is Raspbian based on the Linux kernel. The cloud computing functions are implemented on a laptop, of which the configuration is as follows: the processor is 2.3 GHz Intel Core i7, the RAM memory is 16 GB and the operating system is Ubuntu 16.04.

\subsection{Experimental Results}
\begin{figure} [ht]
\begin{center}
\begin{tabular}{c}
\includegraphics[height=7cm]{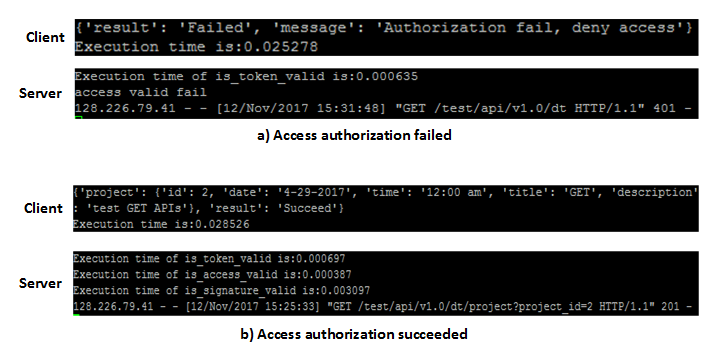}
\end{tabular}
\end{center}
\caption[example] { \label{fig:6-Fig6_CapAC_result} Experimental results of FedCAC system.}
\end{figure}

In the test scenario, one Raspberry Pi 3 device worked as the client while another Raspberry Pi 3 device worked as the service provider. Figure \ref{fig:6-Fig6_CapAC_result} shows the test results.

Given the authorization process shown in Fig. \ref{fig:5-Fig5_CapAC_processing}, when any of the steps in the evaluation task fail, the access authorization process will immediately be aborted instead of continuing to carry out all authorization stages. As shown by Fig. \ref{fig:6-Fig6_CapAC_result}(a), the server aborted authorization process due to fails to verify granted actions or conditional constricts that are in the access right list. Therefore, the client node received a \textit{deny access notification} from the server and cannot read the data it intended to request. In contrast, Fig. \ref{fig:6-Fig6_CapAC_result}(b) shows a successful data request example. The whole authorization process is accomplished at the server side without any error. As a result, the client successfully retrieves the data from the service provider.

\subsection{Performance Evaluation}
To measure the general overhead of the proposed FedCAC scheme on network communication, 50 test runs have been carried out based on the proposed test scenario, where the client sends a data query request to the server for access permission. This test scenario is based on an assumption that the subject has a valid capability token when it performs the action. Therefore, all steps of authorization validation must be processed on the server side so that the maximum latency value is computed.

\begin{figure} [ht]
\begin{center}
\begin{tabular}{c}
\includegraphics[height=6.5cm]{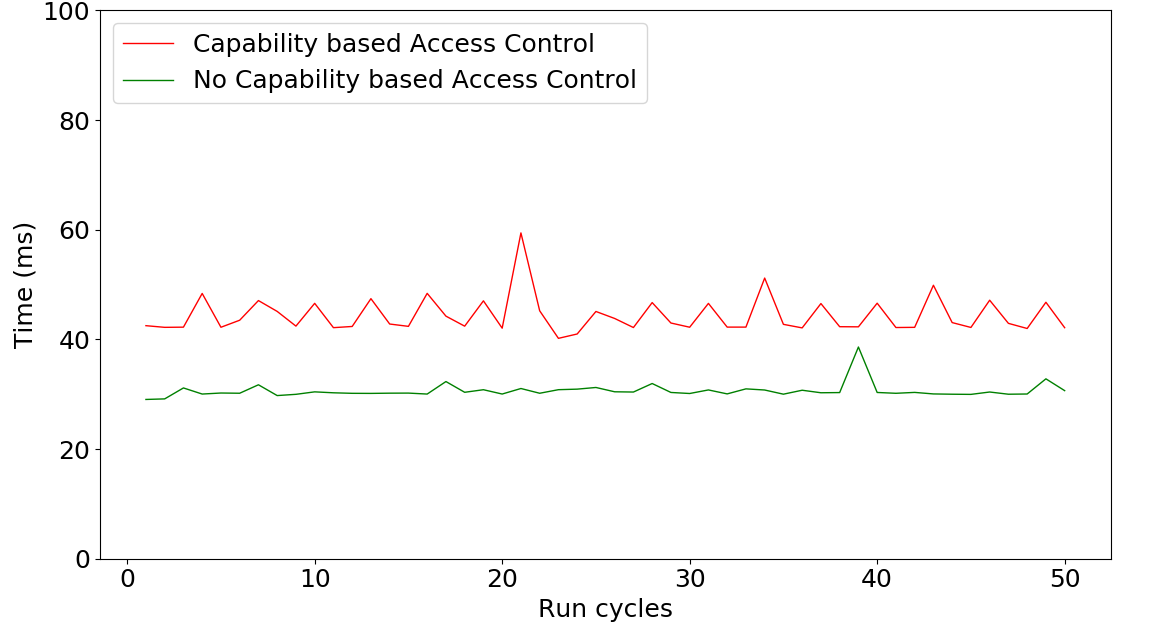}
\end{tabular}
\end{center}
\caption[example] { \label{fig:7-CapACVsNoCapAC} Network latency of our FedCAC.}
\end{figure}

\begin{figure} [ht]
\begin{center}
\begin{tabular}{c}
\includegraphics[height=6cm]{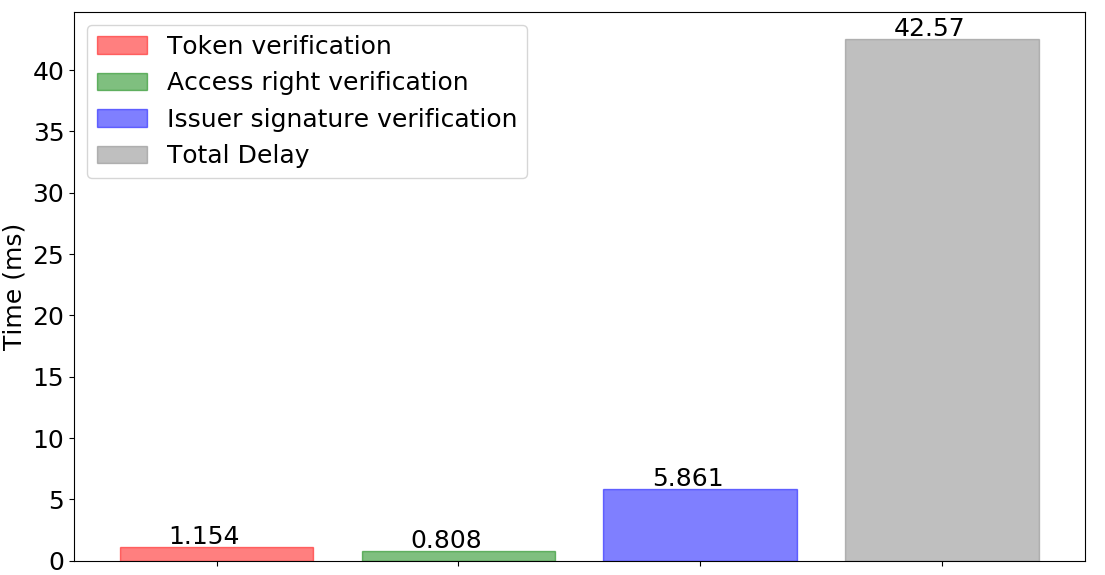}
\end{tabular}
\end{center}
\caption[example] { \label{fig:8-CapAc_exec_time} Execution time of each individual stage of the FedCAC.}
\end{figure}

\begin{itemize}
\item[1)] \emph{General overhead}: Through simulating a regular transaction, we can measure how long it took for the client to send a request and retrieve the data from the server. Figure \ref{fig:7-CapACVsNoCapAC} shows overall network latency incurred by comparing the execution time of the FedCAC to a benchmark without capability based access control enforcement. With the assumption that identity authentication and trust communication channel has been established, the benchmark execution takes an average of 31 ms for searching data versus the FedCAC consumes on average of 42 ms. It means that the capability based access control operation only introduces about 11 ms extra latency in communication. In our test scenario, the client and server nodes are deployed in same network domain, such there is less impact from the round-trip delay time (RTT) than in the inter-domain scenarios. Therefore, our assess is accurate with minimal latency.

\item[2)] \emph{Execution time}: According to the results shown in Fig. \ref{fig:8-CapAc_exec_time}, the average time required for the FedCAC operation of retrieving data from the client to server is 42.57 ms. It includes the RTT, time for parsing JSON data from request and access right validation. Since the authorization process includes token verification, access right verification and issuer signature verification, the average time of authorization process is about 7.823 ms. Given the result that latency incurred by implementing authorization is 11 ms, we can calculate that average time of parsing JSON data from request is about 3.2 ms. As the most computational intensive step, the execution time of the issuer signature verification counted for almost 75\% of whole authorization process time. That is why signature verification is the last step of the authorization process to optimize the access right validation process.
\end{itemize}

As we can see from the experimental results, our proposed FedCAC introduced a smaller overhead both over the network and local device. To measure general network latency of inter domain, httping is executed on same testbed to simulate a regular transaction, like connects, sends a request and retrieves the reply. Compared with calculated average network latency that is about 300 ms, the trade-off in our proposed FedCAC is acceptable for network environments by only incurring 11 ms latency (no more than 4\%). In addition, test scenarios are based on Raspberry Pi which is a type of simple board computer with limited computation and memory, where the performance will be improved through implementing FedCAC on more powerful devices, like smart phone. 

\subsection{Discussions}
The experimental results demonstrate that our proposed FedCAC strategy is effective and efficient to protecting the IoT devices from an unauthorized access request. Compared to centralized CapAC access model, our FedCAC scheme has the following advantages:

\begin{itemize}
\item \emph{Load balance}: The delegation mechanism allows the load of centralized PDC be distributed to separate domain coordinators, such that the bottleneck effect of PDC is mitigated and the risk of malfunction resulting from centralized system is reduced. Even in the worst case when the PDC crash for a short period of time, a large number of domain coordinators still work normally on behalf of the PDC to provide services.

\item \emph{Decentralized authorization}: Since the coordinator (delegatee) delegates part of the privileges from centralized PDC (delegator), it could define domain-specific access authorization policy, which is meaningful to the scalable, heterogeneous and dynamic IoT-based applications;

\item \emph{Fine granularity}: Enforcing access right validation on local service providers empowers those smart devices to decide whether or not to grant access to certain services according to local environmental conditions. Fine-grained access control with lease privilege access principle prevents privilege escalation, even if attacker steals capability token;

\item \emph{Lightweight}: Compared to XML-based language for access control, such as XACML, JSON is a lightweight technology that is suitable for resource constrained platforms. Given the experimental results, our JSON based capability token structure introduces small overhead on the general performance.
\end{itemize}

\section{Conclusions}
\label{sec:con}
In this paper, we proposed a federated capability-based access control (FedCAC) scheme to tackle the challenges in access control strategies for IoTs. Leveraging the fog computing paradigm, a concept-proof prototype has been built to validate the FedCAC. In the  FedCAC system, a laptop serves as cloud computing server to function as a centralized PDC. A Raspberry PI device works as a fog computing node to take responsibility of coordinator, or plays the role of edge computing to provide IoT-based service. Experimental studies have been conducted and the results show that the proposed scheme can efficiently and effectively to enforce access control authorization and validation in a distributed IoT network. This work has demonstrated that our proposed FCapAC framework is a reasonable approach to provide a scalable, fine-grained and lightweight access control in IoT network.

For future research work, our on-going efforts include the following tasks:

\begin{itemize}
\item[1)] \emph{Intelligence on edge networks}: Although the delegation mechanism in our FedCAC migrates certain intelligence from centralized cloud server to a near-site coordinator, the power of policy decision making and identity management is exclusively located on cloud center. IoT networks need a new access control framework that allows intelligence to be diffused among large number of distributed edge devices. We are exploring to build a distributed access control system in which devices are the master to control their own resources instead of being supervised by a centralized authority.

\item[2)] \emph{Network decentralization}: We are also investigating a feasible decentralized authentication and authorization scheme in trustless networks. Owning to key characteristics, such as decentralization, anonymity and openness, blockchain becomes a promising technology worked in a decentralized environment. Exploring blockchain technology in access control strategy will be helpful to solve the dilemma of centralized and decentralized authentication and access control management challenges in IoT field.

\item[3)] \emph{Data heterogeneity}: Combination of IoT with data from other sources such as text, multimedia images, or video such as Wide Area Motion Imagery (WAMI) \cite{wu2017container} as well as that from models including social, cultural, or terrain; respectively.
\end{itemize}

\bibliography{report} 
\bibliographystyle{spiebib} 

\end{document}